\documentclass[12pt]{article}
\usepackage{amsmath}
\usepackage{euscript}

\usepackage{pstricks,epsfig,amssymb,axodraw}
\usepackage{cite}  
\begin{document}

\def\lb{\nextline}
 
\def\Order#1{{\cal O}($#1$)}
\def\Kuhn{K\"uhn}
\def\alphapi{\Bigl({\alpha\over\pi}\Bigr)}
\def\sovem{{s\over m^2_e}}
\def\Born{{\rm Born}}
\def\nubar{\bar{\nu}}
\def\nubarnu{\bar{\nu}\nu}
\def\nubart{\bar{\nu}_\tau}
\def\sstrut{$\strut\atop\strut$}

  \def\PL #1 #2 #3 {Phys. Lett. {\bf#1}           (#3)  #2}
  \def\NP #1 #2 #3 {Nucl. Phys. {\bf#1}           (#3)  #2}
  \def\PR #1 #2 #3 {Phys. Rev. {\bf#1}            (#3)  #2}
  \def\PP #1 #2 #3 {Phys. Rep. {\bf#1}            (#3)  #2}
  \def\PRL #1 #2 #3 {Phys. Rev. Lett. {\bf#1}     (#3)  #2}
  \def\CPC #1 #2 #3 {Comp. Phys. Commun. {\bf#1}  (#3)  #2}
  \def\ANN #1 #2 #3 {Annals of Phys. {\bf#1}      (#3)  #2}
  \def\APP #1 #2 #3 {Acta Phys. Pol. {\bf#1}      (#3)  #2}
  \def\ZP  #1 #2 #3 {Z. Phys. {\bf#1}             (#3)  #2}

\def\uncatcodespecials{\def\do##1{\catcode`##1=12 }\dospecials}
\def\setupverbatim{\tt
  \def\par{\leavevmode\endgraf} \catcode`\`=\active
  \obeylines \uncatcodespecials \obeyspaces \parindent=5mm \parskip=0pt}
{\obeyspaces\global\let =\ } 
{\catcode`\`=\active \gdef`{\relax\lq}}
\def\beginverbatim{\par\begingroup\setupverbatim\doverbatim}
{\catcode`\|=0 \catcode`\\=12 
  |obeylines|gdef|doverbatim^^M#1\endverbatim{#1|endgroup}}
\def\Was{\hbox{W\c as\;}}
\def\M{\hbox{\cal M}}
\def\lips{\hbox{Lips}}
\def\Im{\hbox{Im}}
\def\GeV{\hbox{GeV}}
\def\Maj{M_{R}}  \def\Gaj{\Gamma_{R}}
\def\beq{\begin{equation}} 
\def\eqiv{\sim}
\def\eeq{\end{equation}} 
\def\eps{\epsilon}
\begin{titlepage}
 
\begin{flushright} CERN-PH-TH/2006-025, \\
{ TTP06-01, IFJPAN-IV-2006-1}
\end{flushright}
 
\vspace{0.5cm}
\begin{center}
{\bf\Large
$\tau$ Decays to Five Mesons in TAUOLA
}\end{center}
 
\begin{center}
   {\bf  Johann H. K\"uhn} \\
   {\em Institut f\"ur Theoretische Teilchenphysik,
   Universit\"at Karlsruhe,\\
   76128 Karlsruhe, Germany\\ }
  and\\ 
  {\bf  Z. W\c{a}s } \\
   {\em  PH-TH Dept.\\
       CERN, 1211 Geneva 23, Switzerland\\
        and \\
        Institute of Nuclear Physics, PAN,
        Krak\'ow, ul. Radzikowskiego 152, Poland}\\
\end{center}
\vspace{1.5cm}
\begin{center}
{\bf   ABSTRACT}
\end{center}
The $\tau$-decay library TAUOLA has gained popularity over the last decade.
However, with the continuously increasing precision of the data, some of its 
functionality has become insufficient. 
One of the requirements is the implementation of decays
into five mesons plus a neutrino with a realistic decay amplitude. This note
describes a step into this direction.
For the $2\pi^-\pi^+2\pi^0$ mode the three decay chains
$\tau^- \to a_1^- \nu \to 
\rho^-(\to \pi^-\pi^0) \omega (\to \pi^-\pi^+\pi^0)  \nu$,
$\tau^- \to a_1^- \nu \to 
a_1^-(\to 2\pi^-\pi^+) f_0 (\to 2\pi^0)  \nu$, and
$\tau^- \to a_1^- \nu \to 
a_1^-(\to \pi^-2\pi^0) f_0 (\to \pi^+\pi-)  \nu$ are introduced
with simple assumptions about the couplings and propagators of the
various resonances. Similar amplitudes (without the
$\rho\omega$ contributions) are adopted for the $\pi^-4\pi^0$
and $3\pi^-2\pi^+$ modes. 

The five-pion amplitude is thus based on a simple model, 
which, however, can be considered as a first realistic  example. 
Phase-space generation includes the possibility of
presampling the $\omega$ and $a_1$ resonances, in one channel only, however.
This is probably sufficient for the time being, both for physics applications
and for tests.

The technical test of the new part of the generator
is performed by  comparing Monte Carlo and analytical results.
To this end a non-realistic, but easy to calculate,
purely scalar amplitude for the decay into five massless pions was used.
 
\vspace{1cm}
 
\vspace{.5cm}
\begin{flushleft}
{  CERN-PH-TH/2006-025, \\ TTP06-01, \; IFJPAN-IV-2006-1 \\
 February, 2006}
\end{flushleft}
 
\end{titlepage}
\vspace{.5cm}
\begin{center}
{\bf 1. INTRODUCTION}
\end{center}
\vspace{.5cm}

Early studies of semileptonic tau decays have concentrated on
final states with few mesons only. The recent advent of high-statistic
samples in experiments at LEP and CESR, and the perspective of 
further increasing event rates at B-meson factories allow and require the
study of relatively rare decay modes and thus of multibody final states.
These configurations are of particular importance for a precise
determination of the semileptonic branching ratio and for an improved
limit on the mass of the tau neutrino. Final states with up to six
pions have been observed up to now.

Complementary to the experimental studies, Monte Carlo simulations are
required to determine  the efficiencies of the detectors.
The generator TAUOLA \cite{Jadach:1990mz,Jezabek:1991qp,Jadach:1993hs} has been specifically designed to simulate a wide 
variety of tau-decay modes and includes spin effects in the case of polarized
tau decays.  
These simulations necessarily include form factors that model
the resonant structure of intermediate hadronic states, such as $\rho$,
$\omega$ or $a_1$ mesons. The combination of Monte Carlo simulation and
experimental studies thus allows us to test the model input and leads to
additional information about hadron physics at low energies.

For a few channels only, and in a limited kinematical
range, the form factors can be predicted from a firm theoretical
basis. In many cases additional input, such as vector dominance and 
phenomenological parametrizations  are required. At present only 
final states with up to four pions can be simulated on the basis of
realistic form factors \cite{Jadach:1993hs}, which have their basis in chiral
Lagrangians and vector-dominance models.
In the present paper we describe an extension
of TAUOLA to five-pion final states. The model for the amplitude is
based on the observation  that the $2\pi^- \pi^+ 2\pi^0$ decay 
is dominated by the $\omega \pi^- \pi^0$ channel, with
only about 20\% left for the remainder, which does not exhibit sharp
resonance structures. The amplitude is therefore constructed to
accommodate this feature. It includes the dominant $W^*\to\omega 2\pi$
amplitude and 
an amplitude of the form $W^*\to a_1 (\to \rho\pi) f_0(\to\pi\pi)$
with various charge assignments. The second amplitude is also used
to describe the two remaining charge combinations, $\pi^-4\pi^0$ and
$3\pi^-2\pi^+$. (The charge-conjugate combinations for the $\tau^+$ decay
are considered in parallel.)
In the narrow-width approximation, important technical tests of the
generator can be performed. 
The details of the new amplitude (effectively the transition matrix
element of the hadronic current, often just denoted ``current'')
 and of the phase-space generation
are described in section 2. Tests of the
program and results for some characteristic  distributions are collected
in section 3. Section 4 contains our summary and conclusions.

\vspace{.5cm}
\begin{center}
{\bf 2.  Hadronic currents and phase-space generation}
\end{center}
\vspace{.5cm}

For each decay mode
the basic ingredients for TAUOLA can be grouped into three parts:
(i) the phase-space generator for a fixed number of particles
in the final state,
(ii) the algorithm for the calculation of the spin-dependent matrix
element from the 
hadronic currents and from the properties of the electroweak
interaction between the $W$ boson, the
$\tau$ lepton and its neutrino, and (iii) the hadronic current per se.

The structure of the program is decscribed in detail in refs. \cite{Jadach:1990mz,Jadach:1993hs} (see also \cite{Golonka:2003xt})
and there is no need to recall it here. The relation between the
hadronic current
and the matrix element remains the same as for the two-, three- and
four-meson case presented before. 

Some comments are required concerning the   five-pion phase-space
generator, although its structure is very similar to the well-documented
case of four pions. The number of generated angles and invariant 
masses is, of course, increased. Since no extensive 
studies of amplitudes with narrow resonances are foreseen for the moment,
the phase-space presampler is only prepared to compensate for
possible sharp peaks in the invariant mass of the five-pion system as a whole
 and in one subsystem of the three-pion  (single
channel) only.  
To improve the efficiency, this mode of generation is merged with the 
flat phase-space subgenerator, again according to the rules
described in detail in \cite{Jadach:1990mz,Jadach:1993hs}. 

The most interesting ingredient necessary for this generator  
is the matrix element of the hadronic current. For the decay into an 
$n$-pion state, it is given by
\beq
J_\mu (q_1, q_2, ..., q_n) \equiv
\langle \pi(q_1), \pi(q_2), ..., \pi(q_n) | J_\mu(0) |0 \rangle,
\eeq
where the same letter $J$ has been used for the operator and its matrix
element. The decay into an odd number of pions
proceeds through the axial part of the current only.

We now construct the amplitude as a sequence of (partly virtual) resonance
decays and transitions, and concentrate, in a first step, on the most
complicated  channel $2\pi^- \pi^+ 2\pi^0$. 
From experiment we observe that the decay through $\omega$ seems to
constitute the dominant mode in this case, with a branching ratio 
Br($\tau\to h^-\omega\pi^0$) = ($4.4\pm0.5)\times 10^{-3}$, to be
compared with Br($\tau\to h^-h^-h^+2\pi^0 (exl. K^0,\omega,\eta$)) = 
($1.1\pm0.4)\times 10^{-3}$ \cite{Eidelman:2004wy}.
The first part
will be implemented in a first step (current A), and the remainder
(current B) afterwards. Normalizations will be appropriately set, as can be seen in Section 3.
From isospin conservation --- and the fact that the total hadronic system has
isospin one, and the $\omega$, however, isospin zero --- it follows that the two
pions  are also in an isospin-one 
configuration, corresponding to a state with $\rho$-meson-like quantum
numbers. We thus have to find a convenient description for the amplitude  
that specifies the transition $W^*(Q)\to \omega\rho$.
Let us denote the polarization and momentum of the $\rho$ meson by $\epsilon_\rho$
and $p_\rho$, and similarly for the $\omega$ meson.
Three possible amplitudes describing the decay of a $1^{++}$
to two $1^{--}$ states can be constructed:

\begin{eqnarray}
F^1_\mu &=& ( \eps_\rho , \eps_\omega , Q, \mu )  \nonumber\\
F^2_\mu &=& ( \eps_\rho , \eps_\omega , p_\rho - p_\omega, \mu )   
        - ( \eps_\rho , \eps_\omega , p_\rho - p_\omega, Q ) \frac{Q_\mu}{Q^2}
\nonumber\\
F^3_\mu &=& ( \eps_\rho , \eps_\omega , p_\rho , p_\omega ) 
    \Bigl( (p_\rho - p_\omega)_\mu - (p_\rho - p_\omega)Q
            \frac{Q_\mu}{Q^2} \Bigr),
\end{eqnarray}
each of which is, of course, multiplied by a function of $Q^2$, where
 $Q=\sum_i q_i$.
 For
simplicity we adopt amplitude $F^1$, which depends on the lowest power
of the
relative momentum $ p_\rho - p_\omega $. Furthermore we multiply this
amplitude by a Breit--Wigner factor 
\beq
c(Q^2) = c_0 BW_a(Q^2) \equiv c_0 \frac{m_a^2}{m_a^2 -Q^2 - i m_a \Gamma_a}
\label{ceq}
\eeq
simulating the $a_1$ enhancement. For the constant $c_0$, which
is determined by the product of  $W^*a_1$- and
$a_1\omega\rho$-coupling we adopt the value $c_0=3$, so as to
reproduce the desired branching ratio of $0.4$\%. 
Furthermore we take $m_a=1.26$~GeV and $\Gamma_a= 0.4$~GeV.
With the symbol $( \eps_\rho , \eps_\omega , Q, \mu )$ we denote 
the totally antisymmetric Levi-Civita symbol, contracted 
with three four-vectors $\eps_\rho$, $\eps_\omega$ (polarization vectors for 
$\rho$ and $\omega$), and $Q$; the last index $\mu$ remains open.

The amplitude for the ``subsequent'' decay of the virtual $\rho$ is given by
\beq
{\cal M}_\rho^\mu =g_{\rho\pi\pi} (q_4 - q_5)^\mu,
\label{ro}
\eeq
which leads to the decay rate
\beq
\Gamma_\rho =\frac{|g_{\rho\pi\pi}|^2}{48\pi}
       \frac{(m_\rho^2 - 4 m_\pi^2)^{3/2}}{m_\rho^2}.
\label{rhoP}
\eeq
To reproduce the input parameters $m_\rho=776$~MeV and
$\Gamma_\rho=150 $~MeV the value $g_{\rho\pi\pi}=6.0$ is adopted.
For the $\omega$ we adopt the corresponding decay chain $\omega\to \pi
\rho$, where
all three $\pi$ and $\rho$ charges contribute with equal weight. We
start from an $\omega$--$\rho$--$\pi$ coupling of the form
\beq
{\cal M}_{\omega\rho\pi}=\frac{1}{2} f_{\omega\rho\pi}
(\eps_\omega,p_\omega,\eps_\rho,q_\pi),
\eeq
where the coupling  $f_{\omega\rho\pi}$ has dimension (mass)${}^{-1}$.  
Including the $\rho$ decay according to eq.~(\ref{ro}), and 
taking the antisymmetric isopin
wave function of the three points into account, which implies an  antisymmetric
momentum wave function, we finally arrive at
\beq
{\cal M}_{\omega3\pi}=\frac{f_{\omega\rho\pi} g_{\rho\pi\pi}}{m_\rho^2}
     (\eps_\omega,q_1,q_2,q_3) (BW_\rho(s_1) + BW_\rho(s_2) + BW_\rho(s_3)),
\eeq
where $q_1$, $q_2$, $q_3$ denote the momenta of $\pi^+$,  $\pi^-$,  $\pi^0$, and 
$s_1=(q_2+q_3)^2$ etc. This leads to the decay rate 
\begin{eqnarray}
\Gamma_\omega&=&\frac{1}{3} \frac{1}{128}\frac{1}{(2\pi)^3}\frac{1}{m_\omega^3}
     \frac{(f_{\omega\rho\pi} g_{\rho\pi\pi})^2}{m_\rho^4}
\int {\rm d}s_1{\rm d}s_2
|BW_\rho(s_1)+BW_\rho(s_2)+BW_\rho(s_3)|^2 \\
&&\times(s_1s_2s_3 - m_\pi^2 (Q^2-m_\pi^2)^2). \nonumber
\label{omega}
\end{eqnarray}
If the sum of the three $\rho$ Breit--Wigners is dropped (i.e. replaced
 by a constant taken to be 1) and the pion masses are set to zero, one finds 
\beq
\Gamma_\omega = \frac{1}{3} \frac{1}{128}\frac{1}{(2\pi)^3}m_\omega^7
     \frac{(f_{\omega\rho\pi} g_{\rho\pi\pi})^2}{120 m_\rho^4}.
\label{rhoO}
\eeq
This  formula will be useful for cross checks of the program.
Whenever numerical values are required, we take 
$f_{\omega\rho\pi}=0.07$ MeV$^{-1}$, that is only the first significant digit,
 $g_{\rho\pi\pi}=6$ was defined earlier, $m_\omega=782$~MeV and  
$\Gamma_\omega=8.5$~MeV.

The full five-pion amplitude is obtained by including the $\rho$ and
$\omega$ propagators
\begin{eqnarray}
J_\mu^A(q_1,q_2,q_3,q_4,q_5)&=&
c_0\frac{f_{\omega\rho\pi} g_{\rho\pi\pi}^2}{m_\rho^4 m_\omega^2}
   BW_a(Q^2) BW_\rho((q_4 + q_5)^2) BW_\omega( (q_1+q_2+q_3)^2) \label{JA} \\
 &&(\mu, q_4-q_5, \alpha, Q) (\alpha,q_1,q_2,q_3) (BW_\rho(s_1) + BW_\rho(s_2) + BW_\rho(s_3)).\nonumber
\end{eqnarray}

The pictorial illustration of this decay amplitude is shown in Fig. 1a.
\begin{figure}[!ht] 
\begin{center}
 \begin{picture}(380,150)(-10,-30)
   \Line(1,70)(46,70)   \Text(-10,74)[lb]{$\tau$}
   \Line(46,70)(100,90)   \Text(90,100)[lb]{$ {\nu} $}
   \Line(46, 70)(66,45)  \Text(62,62)[lt]{$a_1 $} 
   \Line(66, 45)(106,45)  \Text(86,37)[lt]{$\omega$}
   \Line(66,45)(56, 15)  \Text(68,30)[lt]{$ \rho$}
   \Line(56,15)(36, -15)  \Text(36,-22)[lt]{$ \pi$}
   \Line(56,15)(80, -15)  \Text(80,-22)[lt]{$ \pi$}
   \Line(106,45)(136, 40)  \Text(146, 40)[lt]{$ \pi$}
   \Line(106,45)(136, 70)  \Text(146, 70)[lt]{$ \pi$}
   \Line(106,45)(136, 10)  \Text(146, 10)[lt]{$ \pi$}
 \Text(-10,100)[lb]{\bf a)}

   \Line(255,70)(300,70)   \Text(244,74)[lb]{$\tau$}
   \Line(300,70)(345,90)   \Text(335,100)[lb]{$ {\nu} $}
   \Line(300,70)(320,45)  \Text(316,62)[lt]{$a_1 $} 

   \Line(320,45)(310,15)  \Text(300,32)[lt]{$a_1 $}
   \Line(320,45)(360,55)  \Text(340,41)[lt]{$f $}
   \Line(360,55)(380, 75)  \Text(390, 75)[lt]{$ \pi$}
   \Line(360,55)(380, 25)  \Text(390, 25)[lt]{$ \pi$}
   \Line(310,15)(340, 5)  \Text(325, -2)[lt]{$\rho $}
   \Line(310,15)(290,-5)  \Text(290, -15)[lt]{$\pi $}
   \Line(340,5)(360,10)  \Text(370, 10)[lt]{$\pi $}
   \Line(340,5)(360,-20)  \Text(370, -20)[lt]{$\pi $}
 \Text(245,100)[lb]{\bf b)}
%
 \end{picture}
\end{center}
\label{Fig1}
\caption{\it Dominant decay amplitude for the decay of $\tau$ into five pions
through an $\omega$ plus a $\rho$ resonance (a) and through an $f_0$ plus
$a_1(\to \rho\pi)$ (b).}
\end{figure}
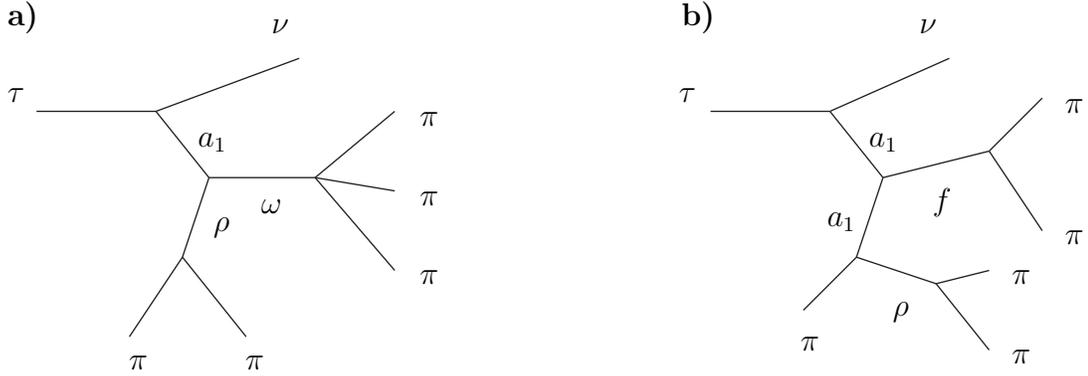
 
This formula is directly applicable to the case of a narrow $\omega$, if we
identify, as stated above, $q_1$, $q_2$, $q_3$ with the momenta of $\pi^+$,
$\pi^-$ and  $\pi^0$ from the $\omega$ decay,  $q_4$ and $q_5$ with the momenta of
the remaining $\pi^-$ and $\pi^0$. Without this approximation the 
symmetrization
between $q_2$ and $q_4$ on the one hand, and $q_3$ and $q_5$ on the other hand has
to be performed 
\begin{eqnarray}
J_\mu^{A;{\rm total}}(q_1,q_2,q_3,q_4,q_5)&=
 &J_\mu^A(q_1,q_2,q_3,q_4,q_5) + J_\mu^A(q_1,q_4,q_3,q_2,q_5)+\nonumber\\
&&J_\mu^A(q_1,q_2,q_5,q_4,q_3) + J_\mu^A(q_1,q_4,q_5,q_2,q_3)\label{JA-sym}
\end{eqnarray}
and the phase space has to be divided by 4 to take the
identical particle statistical factor into account. 

For a test of the proper implementation of the matrix element
(eq. (\ref{JA})), the narrow-width approximation for $a_1$, $\rho$ and $\omega$ is employed. (For
this test $(m_\rho + m_\omega)<m_a$ must be assumed.) In this approximation 
the rate derived from the current $J^A$ can be integrated analytically. 
(Again the sum of three $\rho$
Breit--Wigner amplitudes in the $\omega$ decay is replaced by 1 and
$m_\pi$ is set to zero for this test.)
\begin{eqnarray}
\Gamma_A/\Gamma_e&=& 
\frac{1}{4}\frac{m_{a_1}^2}{m_\tau^2}\Bigl(1-\frac{m_{a_1}^2}{m_\tau^2}\Bigr)^2
\Bigl(1+2\frac{m_{a_1}^2}{m_\tau^2}\Bigr)c_0^2 \frac{m_{a_1}\pi}{\Gamma_{a_1}}
f(m_{a_1}^2,m_\rho^2,m_\omega^2) R
\nonumber \\
f(m_{a_1}^2,m_\rho^2,m_\omega^2)&\equiv&
2\frac{2p}{m_{a_1}}\Bigl(\frac{p^2}{m_\rho^2}+\frac{p^2}{m_\omega^2} + 3\Bigr)
\nonumber \\
p&\equiv&\frac{\lambda^{1/2}(m_{a_1}^2,m_\rho^2,m_\omega^2)}{2m_{a_1}}.
\label{testA}
\end{eqnarray}
The factor 
$R= \frac{\Gamma_\rho^p}{\Gamma_\rho}\frac{\Gamma_\omega^p}{\Gamma_\omega}$
consists of the product of the partial widths for $\rho$ and $\omega$ as given
by eqs.\ (\ref{rhoP}) and  (\ref{rhoO}), divided by total widths used as 
numerical inputs in the Breit--Wigner amplitudes, and the decay rate is
normalized to $\Gamma_e\equiv \Gamma(\tau\to e \bar\nu_e\nu_\tau)$. 

For a test of the generator, in particular of the phase-space
integration, the following, totally  unphysical form of the current
\begin{eqnarray}
J^C_\mu &=& c_1\,\, Q_\mu \nonumber\\
c_1 &=& \frac{1}{m_\tau^3} 4! (4\pi)^3\sqrt{20}
\label{curC}
\end{eqnarray}
was used, with the pion mass again set to zero\footnote{%
As an alternative, in principle we could also use
$c_2= 4! (4\pi)^3 / Q^3$ and obtain the same result. 
In such a case, numerical-stability problems could appear 
as, for masless pions, $\frac{1}{Q^3}$ may approach integrable 
infinity within the allowed phase space.}. 
The analytical result,
\begin{equation}
\Gamma^C/\Gamma_e= \cos^2\theta_{Cabbibo}=0.950625,
\label{testC}
\end{equation}
is well reproduced by the generator. The numerical results of the second test, 
based on the narrow-width approximation, of eq. (\ref{testA}), will be
discussed below.

The decay mode into $\omega \pi \pi $ is obviously only possible for the 
$2\pi^0 2 \pi^- \pi^+$ final state.  In contrast the amplitude
introduced in the following will contribute to all three charge combinations
$\pi^-4\pi^0$, $2\pi^-2\pi^0\pi^+$ and $3\pi^-2\pi^+$. 
For this second amplitude we use the transition of the virtual $W$ into an
$a_1$, with subsequent transition into
$a_1$ plus two pions in an isospin zero, angular-momentum zero
configuration, parametrized by the broad $f_0$ resonance. 
For the $a_1a_1f_0$ coupling the simplest Lorentz structure with minimal
momentum dependence has been adopted.
The $a_1$ subsequently decays into a $\rho$ meson plus a pion, with equal 
amplitude for the two charge modes 
$\pi^-2\pi^0$ and $\pi^+2\pi^-$, similar to the parametrization of the
tau decay into three pions.

This leads to the following current:
\begin{eqnarray}
&&J_\mu^B(p_1,p_2,p_3,p_4,p_5) \equiv \;\;\;\;
\frac{c}{m_a^4 m_{f}^2
m_\rho^2}f_{aaf}f_{f\pi\pi}g_{a\rho\pi}g_{\rho\pi\pi}  \times 
\nonumber\\
&& BW_a((p_1+p_2+p_3+p_4+p_5)^2) 
BW_a((p_1+p_2+p_3)^2)BW_{f_2}((p_4+p_5)^2)\label{JB}\nonumber  \times \\
&&\Bigl[ \Bigl( \frac{Q_\mu Q_\nu}{Q^2} -g_{\mu\nu}      \Bigr)
\Bigl(\frac{p_2(p_1-p_3)}{(p_1+p_2+p_3)^2}(p_1+p_2+p_3)^\nu-(p_1-p_3)^\nu\Bigr)
BW_\rho((p_1+p_3)^2) 
\nonumber\\
&& \; \; + (1 \leftrightarrow 2)\Bigr],
\end{eqnarray}
with the momentum assignments
$a_1(\to \pi_-(p_1) \pi_-(p_2) \pi_+(p_3)) + f(\to \pi_0(p_4)\pi_0(p_5))$.
The pictorial illustration is shown in Fig. 1b.
(In the program, we have used:  $m_{f}=0.8$ GeV, $\Gamma_{f}=0.6$ GeV,          
$G_{a\rho\pi}=6$, $f_{aaf}=4$,  $f_{f\pi\pi}=5$ and      $c=4$.)
The last constant was introduced to normalize the branching ratio for
this channel to 0.11\%. 
The amplitude is, by construction, symmetric under the exchange $p_1$
vs. $p_2$ and $p_4$ vs. $p_5$, as requested from Bose symmetry for the
$f_0$ and $a_1$ decays. Alternatively, the two $\pi^0$ may originate
from the $a_1$ with the $f_0$ then decaying into $\pi^-\pi^+$.
In this case the symmetrization with respect to the momenta of the
$\pi^-$ has to be performed explicitly.
For consistency we have to adopt the same momentum assignment as before:
$\pi^+(q_1)\pi^-(q_2)\pi^0(q_3)\pi^-(q_4)\pi^0(q_5)$. The properly
symmetrized amplitude thus reads:
\beq
J_\mu^{B;{\rm total}}(q_1,q_2,q_3,q_4,q_5)=
 J_\mu^B(q_2,q_4,q_1,q_3,q_5) + J_\mu^B(q_3,q_5,q_2,q_1,q_4) +
J_\mu^B(q_3,q_5,q_4,q_1,q_2). \label{JB-sym}
\eeq

We include the same  statistical factor of $\frac{1}{4}$ as before 
into the normalization of the phase space.
The relative rate
$[ f(+-) + a(00)] : [f(00) + a(-+)] = 2:1$ is recovered
in the narrow width approximation for $a_1$ and $f_0$.
At present there is no analytical 
benchmark for the overall normalization available for this channel.

The extension of this model to the description of the remaining charge
configurations is straightforward.
 Let us start with $\pi^-4\pi^0$ and
adopt the following momentum assignment: 
$\pi^-(q_3)\pi^0(q_1)\pi^0(q_2)\pi^0(q_4)\pi^0(q_5)$.
Using the definition (\ref{JB}) for $J^{B}$ as in eq. (\ref{JB-sym}), the amplitude has to
be symmetrized with respect to the momenta of $\pi^0$. Since $J^{B}$ is already
symmetric with respect to the first two and the last two momenta, only 6
out of the $4!$ permutations have to be considered. This leads to the current
\begin{eqnarray}
J_\mu^{00-00}(q_1,q_2,q_3,q_4,q_5)&=
 &J_\mu^{B}(q_1,q_2,q_3,q_4,q_5) + J_\mu^{B}(q_5,q_2,q_3,q_4,q_1)+
                           J_\mu^{B}(q_2,q_4,q_3,q_1,q_5)\nonumber\\
&+&J_\mu^{B}(q_1,q_4,q_3,q_2,q_5) + J_\mu^{B}(q_1,q_5,q_3,q_4,q_2)+
                J_\mu^{B}(q_4,q_5,q_3,q_1,q_2).\nonumber\\ \label{JBa}
\end{eqnarray}
For the evaluation of the rate the statistical factor of  $1/4!$ must be
included.

In complete analogy we obtain for the $3\pi^-2\pi^0$ mode
\begin{eqnarray}
J_\mu^{--++-}(q_1,q_2,q_3,q_4,q_5)&=
 &J_\mu^{B}(q_1,q_2,q_3,q_4,q_5) + J_\mu^{B}(q_5,q_2,q_3,q_4,q_1)+
                           J_\mu^{B}(q_1,q_5,q_3,q_4,q_2)\nonumber\\
&+&J_\mu^{B}(q_1,q_2,q_4,q_3,q_5) + J_\mu^{B}(q_5,q_2,q_4,q_3,q_1)+
                    J_\mu^{B}(q_1,q_5,q_4,q_3,q_2),\nonumber\\ \label{JBb}
\end{eqnarray}
where the momentum assignment 
$\pi^-(q_1)\pi^-(q_2)\pi^+(q_3)\pi^+(q_4)\pi^-(q_5)$ has been adopted
and a statistical factor $1/2!3!=1/12$ is used in the evaluation of the rate.

For the ``non-$\omega$'' decays and in the narrow-width approximation,
this ansatz predicts the following abundances of the subchannels:
\beq
f(00)a(--+) : f(-+)a(00-) : f(00)a(00-) : f(-+)a(--+) =  
    1       :     2       :      1      :     2
\eeq
and the following relative rates
\beq
2\pi^02\pi^-\pi^+  :  \pi^- 4\pi^0  :  3\pi^- 2\pi^+ = 3:1:2\; . \label{ratios}
\eeq
Note that this prediction is specific to the resonance and isospin
structure of the model. Once more experimental information on the five-pion 
channel will be available, more elaborate possibilities can be
considered.

\vspace{.5cm}
\begin{center}
{\bf 3. Results from the Monte Carlo Program}
\end{center}
\vspace{.5cm}

The new version of TAUOLA includes new channels, numbered 31 to 35, 
which will be discussed in turn. Channels 31, 32 and 33 refer to 
$2\pi^02\pi^-\pi^+$, channel 34 to  $\pi^- 4\pi^0$ and channel 35 to 
$3\pi^- 2\pi^+$. Channels 31 and 32 are implemented for tests, channels
33, 34 and 35 for physics simulations.

Let us start with channel 31, which  is based on current $J^A$, without
symmetrization, as defined in eq. (\ref{JA}). (The order of the pions generated
by TAUOLA is ($--+\,0\,0$). This  ordering is necessary for the test  and the 
requirements of the  presampler running  in the  narrow-resonance  
mode, which  is only
partly optimized. The result, based on the default parameter values listed
after eq. (\ref{ceq}), is  given in  the first  line of  Table  1. This
channel is then used to  test the program against eq. (\ref{testA}), which
is valid  in the  narrow-width approximation  for $\rho$,  $\omega$, and
$a$, using  massless pions. This  simulation was quite demanding  on the
phase-space   generator.  A   precision   of   about   1\%  only   was
reached.  Because of  the  complex structure  of  the resonances  $a_1$,
$\rho$  and  $\omega$  it  was   impossible  to  generate  events 
efficiently.  The $\rho$  Breit--Wigner was  only 
generated from a flat  distribution.  The
variance of the  generated raw sample was deteriorating  quite fast with
decreasing  width,  not only  rendering  the  generator  slow, but  also
risking to arrive at  numerically unstable results.  We nonetheless
completed this test  for several choices of masses  and widths of $a_1$,
$\rho$ and $\omega$. A typical result is listed in Table 1,
line 4, where $\Gamma_\omega=\Gamma_{a}=1$~MeV,
$\Gamma_\rho=5$~MeV, $m_\rho=373$~MeV, was adopted for the parameters,
  and other parameters were left at their default values,
in particular $m_\omega=782$~MeV  and
$m_a=1260$~MeV was kept. 
The remaining 2.5\%
difference between Monte Carlo and analytical calulation 
is  due to contributions from the tails, which  remain large, even
for  the  extremely  narrow   widths  adopted  in  this  example.
   This
illustrates   the   importance   of   non-resonant   contributions   and
interferences, in particular for realistic values of
the parameters.
 
Current $J^C$, which also serves to test the generator, is implemented
in channel 32. In the case of massless pions, eq. (\ref{testC}) is
reproduced with a precision better than 0.1\% (see line 5 of Table 1).
This result is stable and independent of the choice of options for our phase-space 
presampler, providing an important test of its correct operation.
The result for $J^C$, taking the pion mass into account, is listed in
line 3.

In the second line of Table 1 the result is listed for current $J^B$, 
as defined in eq. (\ref{JB}), again without symmetrization.

The results for physical input parameters, and with properly symmetrized
amplitudes, are listed in Table 2, column 4. The first three lines refer 
again to $2\pi^02\pi^-\pi^+$, as implemented in channel 33, and are based on 
eqs. (\ref{JA-sym}) and (\ref{JB-sym}). For the order of the pion momenta in
the program the convention\footnote{Note the difference in order with respect
to  Section 2 and channel 31.}
 ($+--00$) has been used. 
The first line is based on
the $\omega\pi\pi$ channel ($J^{A;{\rm total}}$) alone, and the symmetrization
evidently increases the result by about +5\%. The second line is based
on $J^{B;{\rm total}}$ alone, and the symmetrization decreases the result by
11\%. Adding coherently $J^{A;{\rm total}}$ and  $J^{B;{\rm total}}$ one arrives at
the result given in line 3. Keeping in mind  our previous experiences
with interferences, it is amazing that the result is
exactly the sum of the two previous entries. This  implies that the two
amplitudes do not lead to interferences in the rate in any significant manner!
But of course, perfect agreement for all significant digits is due to 
statistical fluctuations.

The charge configurations $\pi^-4\pi^0$ and $3\pi^-2\pi^+$ 
(with the momentum ordered ($-0\,0\,0\,0$) and ($-++--$) in the TAUOLA output,
again differently as in Section 2)
correspond to channels 34 and 35 respectively. 
The results are based on the amplitudes given in 
eqs. (\ref{JBa}) and (\ref{JBb}) and the numerical predictions are listed 
in lines 4 and 5. 
These numbers are contrasted with the experimental results
listed in Table 2, column 5. The result for $2\pi^-\pi^+2\pi^0$ 
as obtained with $J^{A;{\rm total}}$ alone is compared with the mode 
$h^-\omega\pi^0$, the result for $J^{B;{\rm total}}$
alone with the mode $h^-h^-h^+\pi^0\pi^0$ (exl. $K^0,\omega,\eta$).
The $\pi^-4\pi^0$ mode is compared with $h^-4\pi^0$ (ex $K^0,\eta$) and
$3\pi^-2\pi^+$ with $h^-h^-h^-h^+h^+$. Let us emphasize again that the
rates for the two submodes of channel 33 are fitted to the data, 
the rates for channels 34 and 35 are then the result of our ansatz.

The relative rates as displayed in lines 2, 5 and 6, column 4, are
seemingly in contradiction with the expectations based on the narrow-width 
approximation for the intermediate states, eq. (\ref{ratios}). We
therefore perform the simulation also in the narrow-width approximation,
with the following parameters: $m_f=0.45$~GeV, $\Gamma_f=10$~MeV.  
For the  $a$ saturating intermediate  3-pion state, we took $m_a=1.16$~GeV and $\Gamma_a=10$~MeV, 
while width and mass of the $a_1$ coupled directly to the $\tau$ lepton were
kept at their standard value; this was also true for  the mass and width of the $\rho$ meson. 
Indeed, in this completely fictitious case,
eq.(\ref{ratios}) is recovered within the statistical error (column 6).

\begin{figure}
\begin{center}
\begin{tabular}{|r|r|r|r|}
\hline
Current &  $\Gamma_X/\Gamma_e$ (TAUOLA) & $\Gamma_X/\Gamma_e$ (analytical) 
           & Comment \\
\hline
A  &0.02404  $\pm$3 $\times 10^{-5}$   & ---                    & without symm.     \\
B  &0.00928  $\pm$9 $\times 10^{-6}$   & ---                    & without symm.     \\
C  &0.10690  $\pm$2 $\times 10^{-4}$   & ---                    & massive pions     \\
\hline
A  & 1.38$\times 10^7 \pm$ 1 $\times 10^{5}$ &  1.41$\times 10^7 $ & massless pions    \\
C  &  0.95030 $\pm$ 3 $\times 10^{-5}$ & 0.950625               & massless pions    \\
\hline
\end{tabular}
\end{center}
Table 1: {\it Test results of the generator for unphysical choices of
parameters; see text. }
\end{figure}

\begin{figure}
\begin{center}
\begin{tabular}{|r|r|r|r|r|r|}
\hline
Channel   & Final state  & Current & $\Gamma_X/\Gamma_e\times 10^3$ 
                                   & $\Gamma_X/\Gamma_e\times 10^3$ 
                                   & $\Gamma_X/\Gamma_e$ \\
          &              &         & (TAUOLA)    &
                                   (experiment \cite{Eidelman:2004wy} )
                                                                    &(narrow width)\\ 
\hline
33& $2\pi^-\pi^+2\pi^0$ &A ; total  &25.30 $\pm$0.1\% & $25\pm3$  & -- \\
33& $2\pi^-\pi^+2\pi^0$ &B ; total  & 6.05 $\pm$0.2\% & $6.2\pm2$ & $2192\pm 22$  \\
33& $2\pi^-\pi^+2\pi^0$ &A+B ; total&31.35 $\pm$0.1\% & $31\pm2$  &  --           \\ 
34&   $\pi^-4\pi^0$     &B ; total  & 9.37 $\pm$0.1\% & $5.5^{+3.4}_{-2.8}$
                                             & $788\pm 4$  \\
35& $3\pi^-2\pi^+$      &B ; total  & 11.03 $\pm$0.1\% & $4.6\pm0.3$ & $1469\pm 7 $  \\
\hline
\end{tabular}\vspace{0.3cm}
\end{center}
Table 2: {\it Test results of the generator for realistic choices of
parameters; see text. \\ }
\end{figure}

Eq. (\ref{ratios}) is only reproduced for very small widths
of the resonances. For realistic parameters, the predictions for ratios 
of branching fractions are significantly affected by interferences, and
this applies even more so for differential distributions. 
This observation has to be taken into account from the
very beginning in the formulation of a realistic current and the construction 
of Monte Carlo programs  where  cascade decay chains must be
implemented.

\vspace{.5cm}

\newpage

\begin{center}
{\bf 4. Summary}
\end{center}
\vspace{.5cm}   In  this   note,  recent   developments  in   TAUOLA  for
$\tau$ decays    into   five    mesons   are    described.     For   the
$2\pi^-\pi^+2\pi^0$ mode,  two amplitudes were  introduced. The  first is
motivated    by   the   experimental    observation   of    a   dominant
$\omega\pi\pi$ channel,  the   second  corresponds  to   a  non-$\omega$
contribution  and   is  parametrized   by  the  transition   through  an
intermediate $a_1+f_0$ combination. This  second amplitude was also used
to predict the two  remaining charge combinations for $\tau$ decays into
five  pions.  Both  amplitudes  are  implemented into  TAUOLA.  A  third
current,   $J^C$,  was  introduced   for  tests   of  the   phase-space
generator. With the help of  these currents basic technical tests of the
algorithm were successfully completed.  No sophisticated fits to the data
were attempted. The model amplitude  can, however, be used as a starting
point for more detailed theoretical and experimental investigations.

We  observed  that  effects  due  to  interferences  between  different
resonant amplitudes  are often large.  This is due to  various threshold
effects and may be specific for the choice of our amplitude. It may 
result, for example, from those various threshold effects or from simplifying 
assumptions adopted for our
currents, such as  the choice of  constant instead of running, final-state
mass-dependent widths.  For  different parametrizations, the influence of
interferences on the total rates could be significantly smaller.

This isssue is of great practical importance for the design of a generator
used in  fits to data during  their analysis, when the  general form for
the currents often has  to be changed several  times.  If the constraint
of zero  interference between  various currents, standing  for different
subresonances,  was imposed  on  the algorithm,  as  was the  case
e.g. with EURODEC~\cite{eurodec,Eberhard:1989ve}, the fitting procedures
would become unnecessarily difficult (even impossible), or might even not
lead to a correct description of the amplitude.  
Therefore, distinct  currents for each final state, which are 
labelled by  stable  decay products, will be  also used in
the future. Resonances
such as  $\rho$, $a_1$,  $\omega$ are can  only be used  as intermediate
states in the amplitudes. 

Finally, let us point out that the extensions for TAUOLA as presented in
this paper are available from the
directory {\tt tauola-BBB} of the standard distribution system for TAUOLA
versions. The usage of the directory {\tt tauola-BBB}
is not documented yet; it is, however, identical
to the one of the {\tt tauola-F} subdirectory as 
explained  in \cite{Golonka:2003xt}. 
The most recent version of the code is available from the web page
\cite{tauolaphotos}. Test results for differential distribution, 
obtained with the help of MC-TESTER \cite{Golonka:2002rz}, 
are available from the web page \cite{test}.

\vspace{.5cm}

{\bf Acknowledgements.}
Supported by grants BMBF/WTZ/POL01/103 and BMBF  No. 05HT4VKA/3.
Supported in part  by the EU grant MTKD-CT-2004-510126, in partnership with 
the CERN Physics Department, and  by the Polish State Committee for Scientific Research 
(KBN) grant 2 P03B 091 27 for the years 2004--2006.
One of us (ZW) acknowledges the hospitality of  TTP Karlsruhe where part of this work was done
and the Polish-French collaboration within IN2P3 through LAPP Annecy.

\end{document}